**Hybrid deliberation: Citizen dialogues in a post-pandemic era**

Weiyu ZHANG

Director, The Civic Tech Lab (www.civictechlab.org)

Associate Professor, Department of Communications and New Media, National University of Singapore

Nov 18, 2022

**Acknowledgement**

This report was commissioned by The Bertelsmann Stiftung (The Bertelsmann Foundation).

**Citation:**

Zhang, W. (2022). Hybrid deliberation: Citizen dialogues in a post-pandemic era. Consultancy report for The Bertelsmann Stiftung (The Bertelsmann Foundation). Accessed at:

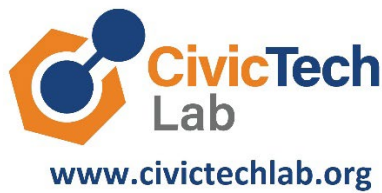

www.civictechlab.org

**Introduction**

    The emergence of information and communication technologies (ICTs) such as the Internet and smart phones has enriched our imagination about citizen participation, especially the dialogue-based participation. Many practical problems associated with the past citizen dialogues are believed to be solvable through ICTs: the technologies allow citizens to overcome physical distance to gather virtually, provide abundant information to equip citizens with the necessary knowledge for policy discussions, enable direct communication between citizens who come from diverse backgrounds, encourage citizens to be inclusive and rational through design features, and give policy makers tools to instantly analyze and learn from citizens' discussions. The motivation of leveraging on ICTs to engage citizens in the increasingly complex policy making process is shared among many polities, regardless of their level of democratization. The Covid-19 pandemic has made this quest for digital, technological solutions to facilitate citizen dialogues all the more acute. When citizens were under lock-down or subject to safe-distancing measures, face-to-face or in-person dialogues became impossible. Digital tools remained as the only way to support citizen dialogues.

    This report first provides a brief review of various forms of dialogue-based participation, e.g., Citizen Assembly, Citizen Lottery, Citizen Jury, Deliberative Polling, and Participatory Budgeting. Challenges associated with these long-lasting practices are identified and hybrid deliberation is proposed as a concept to address the challenges. The report then analyzes six leading examples of digital or hybrid formats of citizen dialogues. Through the comparison of the cases, the report concludes about the hurdles/risks, success factors/opportunities, and best practices for a complementary use of digital and analogue participation formats. Hybrid deliberation is proposed to be the future direction for dialogue-based participation that involves masses and generates high-quality outcomes.

**Background**

    Dialogue-based citizen participation has long history and can be found in many countries. From German tea house to American town halls, the evolution of democracy is indeed a progress towards including more citizens to participate in collective decision making in more equal and effective ways. The 21st century has witnessed a wave of deliberation-based participation, which emphasizes rationality in the dialogue style and fairness in the dialogue procedure (Zhang & Chang, 2014). Practices that embrace the deliberation model have mushroomed around the world, taking different names but implementing similar measures: Citizen Assembly, Citizen Lottery, Citizen Jury, Deliberative Polling, and Participatory Budgeting. Participatory Budgeting (PB), for instance, is a widely practiced deliberation method to engage citizens in budget planning.  Another example is Citizen Jury, a deliberation method rooted in the institution of jury duties. Deliberative Polling combines random sampling of survey respondents, educational materials and talks, and interpersonal (either online or offline) deliberation. The development of ICTs have both enabled deliberation practitioners to move the existing participation mechanisms online and puzzled them with new challenges such as incivility (Zhang, 2005) and inequalities

(Zhang, 2010). This effort has been led by both technologists and academics (see a review of this history in Zhang et al., 2022) The following five cases represent a diverse range of attempts to take advantage of the ICTs while avoid their pitfalls to facilitate citizen-based dialogues (see Table 1 for basic information).

**Case Study**

**Table 1. Comparisons of the five cases.**

|  | ODSG | COLLAGREE – D-Agree | ConsiderIt | Decidim Barcelona | Global Assembly |
|---|---|---|---|---|---|
| geographical scope | Singapore | Japan | US | Spain | International |
| initiator | Academics + policy makers | Academics | Academics-> commercial? | Academics + NGOers | Academics + NGOs |
| history | A follow-up to the SG gov-led Our Singapore Conversation | Deliberative Polling first practiced in 2009 | Following Reflect! and OpinionSpace as early tools | Following Decide Madrid and its Consul platform | A global version of Citizens' Assembly, a practice that can be dated back to the 1970s. |
| running time | 2 months | 2 – 4 weeks | 1.5 months | 2 months | 11 weeks |
| topic | Population policies | City planning | Election reform | drafting the Municipal Action Plan (PAM) | The climate and ecological crisis |
| method | Random sample of SG population + training + asynchronous online discussion + voting | Self-selection of citizens + invited celebrity users + asynchronous online discussion | Self-selection of citizens through newspapers+ school students and club members + online pro/con listing | Self-selection of citizens + asynchronous online discussion + voting | Lottery to choose participants + learning materials + professionally made documentary + synchronous online meetings |
| involved technology | Discussion forum with argumentation input | Discussion forum with tools to support both | Argumentation platform that asks for input on pros | A combination of scheduling | Advanced sampling of a global population + |

| | interface and human-moderated argument map | facilitators and participants to understand discussion content | and cons, and visualize the distribution of opinions | system (to schedule offline meetings) and discussion forum | offline local deliberation + online global deliberation |
|---|---|---|---|---|---|
| outreach | ~2000 citizens | ~800 citizens | ~8000 citizens | ~25,000 citizens | ~100 citizens |
| policy impact | Unclear – only presentations to policy makers | Unknown – not reported in any publications | Unknown – not reported in any publications | 71% of citizen proposals accepted and included in PAM through over 1,600 initiatives | The Core Assembly presented their key messages at COP26, and a declaration was delivered to world leaders. |
| participants (numbers, type) | ~2000 ordinary citizens took the surveys, ~500 ordinary citizens joined the online discussion, ~20 policy makers involved | ~800 ordinary citizens registered at the platform, 3 celebrity users involved in the 2016-2017 instance | ~8000 ordinary citizens visited the website, ~500 ordinary citizens registered at the website | ~25,000 citizens signed up on the platform, 10,860 proposals submitted, 410 meetings held and over 160,000 votes collected | ~100 citizens in the Core Assembly |
| weblink to further information | https://civictechlab.org/projects/online-deliberation | https://d-agree.com/site/en/functions/ | https://consider.it/ | https://docs.decidim.org/en/understand/about | https://globalassembly.org/about-2 |

*Case 1: Online Deliberation SG from Singapore*

Engaging citizens in policy making in Singapore can be dated back to 1985 (Rodan, 2021), when the Feedback Unit (FU) set up. The unit was later revamped in 2006 to become Reaching Everyone @ Home for Active Citizenry (REACH). Government-led committees held consultations with stakeholders regularly. In the 1990s, The Next Lap and Singapore 21 were two large-scale consultations that involved thousands of Singaporeans (Soon & Liang, 2021). Remaking Singapore Committee launched a public consultation program with around 10,000 Singaporeans in 2002. One of the recent nation-wide dialogue-based program was Our Singapore Conversation in 2012. This program included almost 50,000 people and 600 small-group dialogues. Following this tradition, the Online Deliberation Singapore (ODSG) program was launched in 2016, aiming to solve a few remaining challenges from previous efforts: (1) previous programs rely on either self-selection or political actors' nomination/invitation to decide who will be included in the dialogue. This results in non-representativeness of the population and their voices. Marginalized groups, especially the ones who don't have time and resource to afford joining the dialogue, are left out. (2) previous programs conduct face-to-face town halls or small group discussions. Face-to-face discussions have lots of advantages but are more likely to reproduce the existing power balance (Chen & Zhang, 2020). Again, marginalized groups that are not equipped with the skills and agency to participate may risk their voices dismissed, even if they manage to join the dialogues.

   ODSG was funded by Singapore's Ministry of Education as an academic research project and a collaboration between Department of Communications and New Media and Institute of Policy Studies, and School of Computing at National University of Singapore. The project investigates citizen participation in policy discussion on a digital deliberation platform to understand how citizens perceive the effectiveness and legitimacy of online deliberation in policymaking, and to develop a platform that will enable policymakers at various levels to understand and incorporate citizen participation. The project also examines various design features of a digital platform that influence the mechanisms of deliberation and ultimately shape participants' perception of deliberatio. The policy issue selected is the Singapore population and the National Population and Talent Development Division (NPTD) worked closely with the project team to narrow the focus to three sub-issues — Fertility, Foreign Workforce and New Immigrants. The multi-phase project follows three steps to complete the online deliberation.

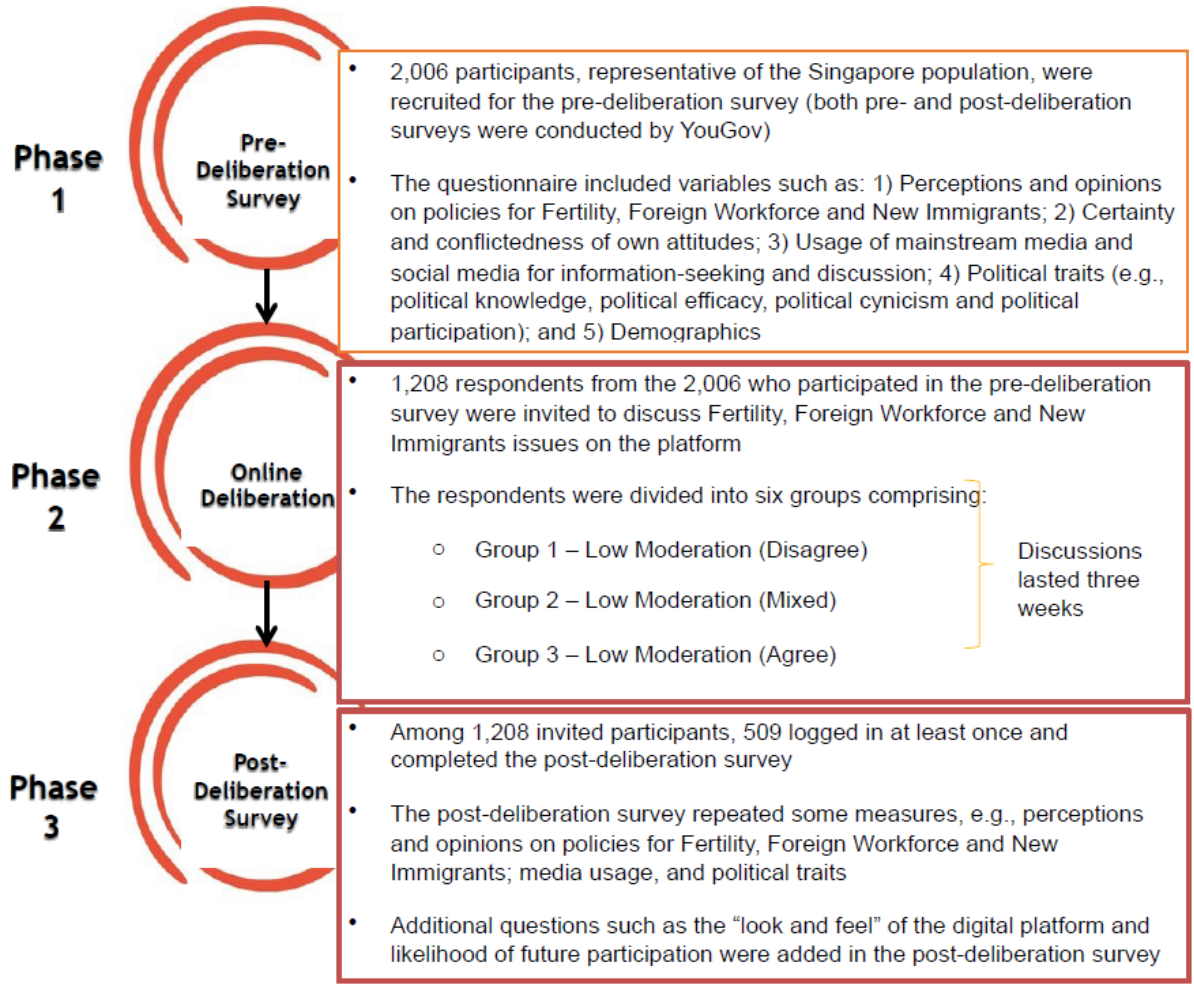

A prominent component of ODSG is its online deliberation platform. The online platform includes various features that engage participants at different phases (a separate interface was designed for policymakers).

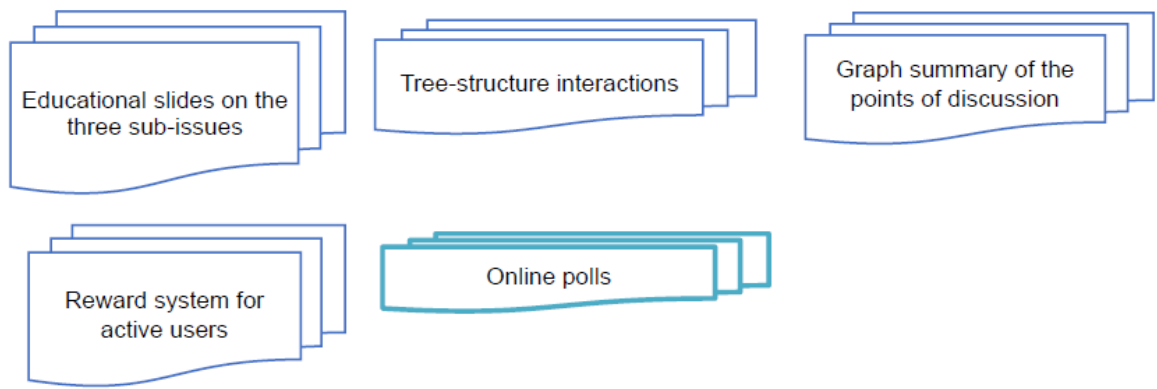

The education slides were designed by **policy makers** and researchers. Citizen users would Go through slides once and answer a set of open-ended questions to prompt their thinking.

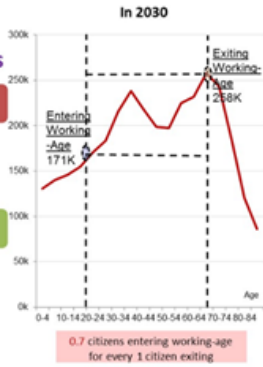

The discussion interface follows a tree-structured system, with options for citizen users to raise an issue, provide an idea, indicate agreement and disagreement. The platform has an additional graph component to allow moderators manually summarize the arguments and present to citizen users for betting understanding of the overall deliberation.

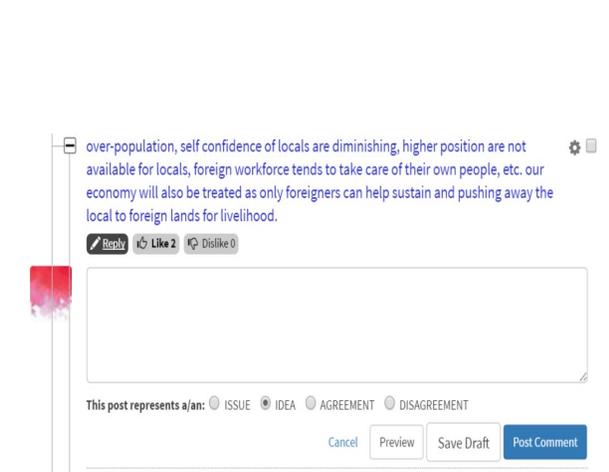
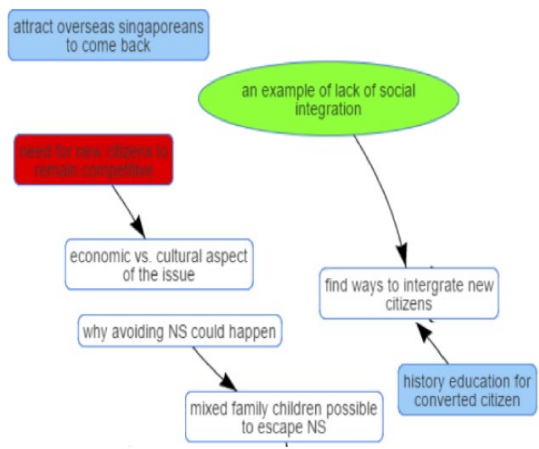

The platform also includes rewards systems to incentivize active participation. The points citizen users received from participation can be transformed into cash rewards. The platform provides a polling function, allowing citizen users to generate concrete recommendations after each round of deliberation. The platform features user guideline and human moderators to ensure that the deliberation is legal and civil.

The results show that eventually 510 users used the platform (Zhang & Perrault, 2019) and 456 of them completed the post-deliberation survey. Citizen users generally liked the deliberation experience. 65-70% of user positively rated the experience, 62-72% liked the interface features, while 54-58% positively rated the system usability. In addition to the platform experience, users were largely positive about the people and views they encountered during the online experience. On a five-point Likert Scale, they agreed that their discussion partners' claims were valid (M = 3.68, SD = .64), the procedure of deliberation was fair (M = 3.71, SD = .67), the decisions generated through online deliberation were legitimate (M = 3.73, SD = .66). On top of these positive evaluation of both the technology and the content of discussion, ODSG was able to move citizen users' opinions in a few directions (Zhang & Soon, 2017). Support for bringing new citizens increased while support for having more foreign workforce decreased.

In summary, ODSG is a citizen dialogue fully mediated through the online technology. It has its strengths in random sampling of the population and half-anonymous participation, both of which could encourage marginalized groups to voice their views. However, ODSG is still not perfect in generating representativeness of the population, with participants skewed towards those who are younger, the more educated and those from the middle-upper class. Another limitation is the lack of visible impact on policy making. Although anonymized data, reports, and presentations were shared to the policy makers, it is unclear whether any of the information was incorporated in later policies.

Key takeaways

| DOs | Do Nots |
|---|---|
| - Use random sampling of the population and supplement with purposive sampling of minority groups | - Don't assume that random sampling leads to representative sample of the population |
| - Use existing technology (e.g., online forums or zoom meetings) to facilitate the dialogue | - Don't use too many cutting-edge technologies as they confuse ordinary citizens |
| - Integrate the outcomes of the online event into the institution of policy making | - Don't just involve policy makers, make them commit |

*Case 2: COLLAGREE from Japan*

Citizen participation methods such as public hearing and public comment were adopted in Japan as early as in the post-WWII era, when the country was re-democratizing itself. Although such participation was common at the local government level, self-selection was a shared problem (Sone, 2021). A first Deliberative Poll (DP) was carried out in 2009 and DP happened at a once-a-year frequency till 2014. The DPs included two national-level deliberation, such as the one on the topic of "Energy and Environment" in 2012. As Sone (2021) pointed out, DP faces several challenges in Japan, such as lack of resources to support face-to-face day-long deliberation events and difficulty to recruit participants. The advent of the Internet seems to be capable of addressing both challenges, lowering down the costs for both DP organizers and participants. COLLAGREE was the first online deliberation practice in Japan, launched in 2013 by a group of computer scientists in collaboration with the Nagoya city. Mayor Takashi Kawamura announced this project in mass media as one actual town meeting to invite citizen participation in discussing the Nagoya Next Generation Total City Planning for 2014-2018 (Ito et al., 2014). After its collaboration with the Nagoya government, COLLAGREE evolved into D-AGREE, a commercial entity that sells its online discussion services to various customers. D-AGREE was launched in 2020 and positioned as a Japanese start-up located in the Nagoya city. In one of its blog posts, D-AGREE said that "We would like to expand the Southeast Asian market with Singapore as an opportunity!"[1]

COLLAGREE was funded by the Nagoya government and supported by expert facilitators from Facilitators Association of Japan. The lead researchers came from Department of Computer Science and Department of Architecture and Design at Nagoya Institute of Technology. The 2013 deliberation ran through a two-week period, and was open for the public to discuss their views about what the city should develop into. The 2013 instance generated 266 registered participants, 1,151 posts, 3,072 visits, and 18,466 views. As the organizers reported, "(t)he total of 1,151 opinions greatly exceeded the 463 opinions obtained by previous real-world town meetings." 2018 (Ito et al., 2014) A second instance occurred during December 12, 2016 to January 9, 2017, to discuss the topic of "Nagoya's appeals". This time, three local celebrities were advertised to join the online deliberation along with the public. The 2016-2017 instance generated 822 registered participants, 1,327 posts and 19,599 views (Nishida et al., 2018). Compared to the 2013 instance, the 2016-2017 instance included more phases, including a "divergence phase", a "convergence phase", and an "evaluation phase".

COLLAGREE was initiated as a government tool to collect feedbacks so its first functions were to facilitate the discussions. The following picture is an illustration of the facilitator functions the platform includes, copied from one publication (Ito et al., 2014).

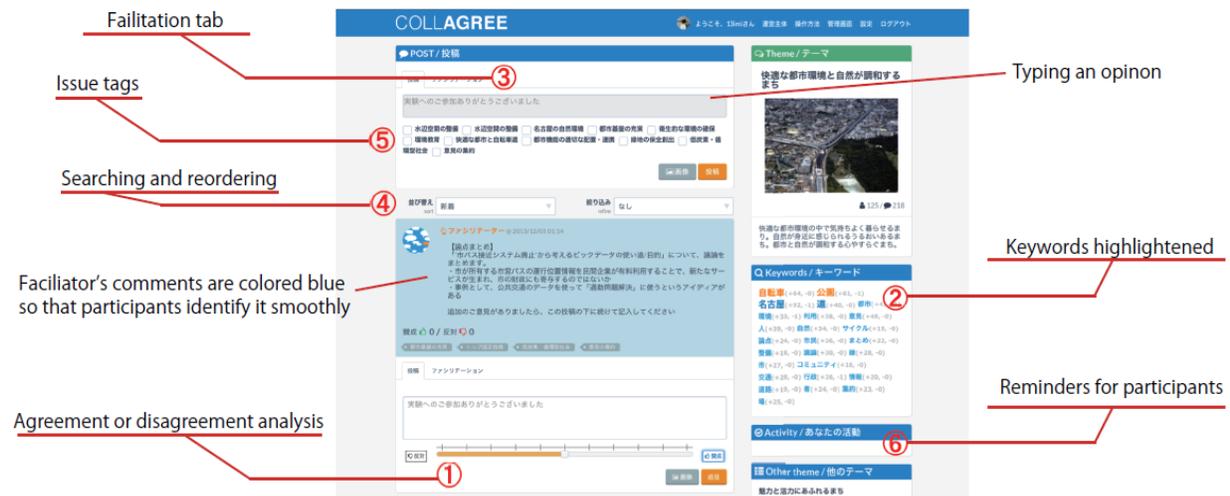

More functions were added to the platform to help participants to discuss in the 2016-2017 instance. The following table summarizes these added functions, copied from one publication (Nishida et al., 2018).

| Function | Explanation |
|---|---|
| (1) Attaching image | Participants can attach images when posting. |
| (2) Issue tags | By adding issue tags to posts, participants can sort and refine searches. |
| (3) Like button | Like button is used to show favor on favorite posts. |
| (4) Discussion points | Participants can get discussion points according to number of posts, replys and likes. Discussion points proposes to active discussion. |
| (5) Keywords highlighted | Keywords highlighted displays words that appeared frequently during discussions. |
| (6) Agreement or disagreement | Participants can express agreement or disagreement when replying to posts. |
| (7) Discussion tree | Discussion tree visualizes flow of a discussion on basis of reply relationships. Participants can grasp hierarchical structure of discussion from this tree visualization. |
| (8) E-mail reminder | E-mail reminder notify participants when there are replies to posts. |

The results (Nishida et al., 2018) show that there are 3 times more male participants than females. More than 50% of the participants came from the 30s and 40s age ranges. Most of participants (75%) did not post anything. The three celebrity users' posts received about 3 times more replies than an average user's post. There were in total 349 responses to voting and 41 to the survey questionnaire. Among the 41 responses, about 80% agreed that "it was good to discuss about Nagoya's appeal on the Web" and 70% agreed that they "want to discuss Nagoya's appeal in the future on the Web".

In summary, COLLAGREE is a public platform that is open to all Nagoya city residents to join. In order to encourage participation, the project engages city mayor and celebrities to call for citizen inputs. However, the scale of participation is still far from ideal. The representativeness of participants is far different from the city population. It is unknown whether the deliberation is able to change citizens' opinions and reach some collective decisions. In the second iteration, celebrity users' views seem to have obtained more support. Only 40% of survey respondents agreed that they are "satisfied with the agreement". It is also unknown whether these agreements had any impacts on the Nagoya city policies.

Key Takeaways

| DOs | Do Nots |
|---|---|
| - Engage professional facilitators to facilitate the online dialogue | - Don't expect that citizens will participate if there is a platform |
| - Government leaders and celebrities help to recruit participants | - Don't over-rely on celebrities who may dominate the dialogue |
| - Do multiple iterations so the dialogue practice can improve | - Don't forget to evaluate citizens' experience using data-driven methods |

*Case 3: ConsiderIt from USA*

The USA was the birthplace of many modern dialogue-based citizen participation mechanisms. From the New England Townhall meetings in the 1600s to Deliberative Polling starting from the turn of the 21$^{st}$ century, American governments and non-government organizations have been active in pushing the deliberation experiments. "The Deliberative Polling methodology uses a randomly selected cohort of citizens, who are provided with a balanced and independently reviewed briefing package, prior to deliberating together about a set of issues or problems." (He & Breen, 2021, p3) Moving deliberation online, the first deliberation platform emerged in the USA under the lead of a group of University of Pennsylvania scholars. The Electronic Dialogue Project was launched during the 2000 American Presidential Campaign (Price & Cappella, 2002) and was followed by a similar iteration on a different public issue, Healthcare Dialogue (Price & Cappella, 2007). As someone who worked for both projects as a research assistant, I see these projects largely an online replication of DP by moving the in-person discussions to synchronous online discussions. New online deliberation experiments provided innovative interfaces/platforms to fit the deliberation goals better. For example, Reflect! modified the comment area on online websites to encourage users to listen to others. OpinionSpace is a mapping tool to allow users of online forums to gauge their opinions' position compared to other users' opinions. ConsiderIt was also chosen because it is a brand new platform that attracted a sufficient number of citizens to participant. The platform was launched by a group of scholars from University of Washington, in collaboration with Seattle City Club, journalists and schools. The platform was introduced during the 2010 Washington State election, open to all Washington voters to share their views, reflect on the tradeoffs, and consider others' perspectives. After the platform being used from launch (9/21/2010) to the election (11/2/2010), ConsiderIt was re-used by a group of German scholars to discuss among their students an issue on "whether Greece should leave the EU monetary union or not." The platform now evolved into a commercial service, similar to the start-up in Japan. However, the co-founders listed on the website are not those who wrote the academic papers. It thus remains more investigation to verify whether the website is indeed linked to the original ConsiderIt.

ConsiderIt lasted for 1.5 months in its 2010 fieldwork and was open for the public to join. Instead of using a random sample of the Washington State population, the project relied on partners such as local newspapers, high schools, colleges, and social organizations to recruit participants (Kriplean et al., 2012). For this first instance, Google Analytics data show the platform "received 12,979 visits from 8,813 unique visitors. Ignoring the 6,082 sessions in which users visited only the homepage, users stayed an average of 10 minutes 39 seconds and visited 6.1 pages." 468 people registered and submitted a position on at least one of 12 measures. "184 con and 160 pro points were written by 147 users. The maximum points contributed by one user was 10. 298 users included these points 2,687 times into 678 unique pro/con lists (503 pro/con lists were left empty)." A second instance was more like an in-class experiment as "36 students participated, 18 per condition. The male to female ratio was 69% versus 31%. Their mean age was 23.1 years (SD = 2.0). They had 3.2 years of college education on average (SD = 1.1)."

(Stiegler & de Jong, 2015) Since the second instance is of little public impact, the review from now on focuses on the first instance.

ConsiderIt Washington was designed to (1) provide 12 election reform measures for users to consider; (2) ask the users to list the pro and cons they think the measure has; (3) by listing, users can both write down their own thoughts and inserting those written by other users. The initial page of the interface looks like the following:

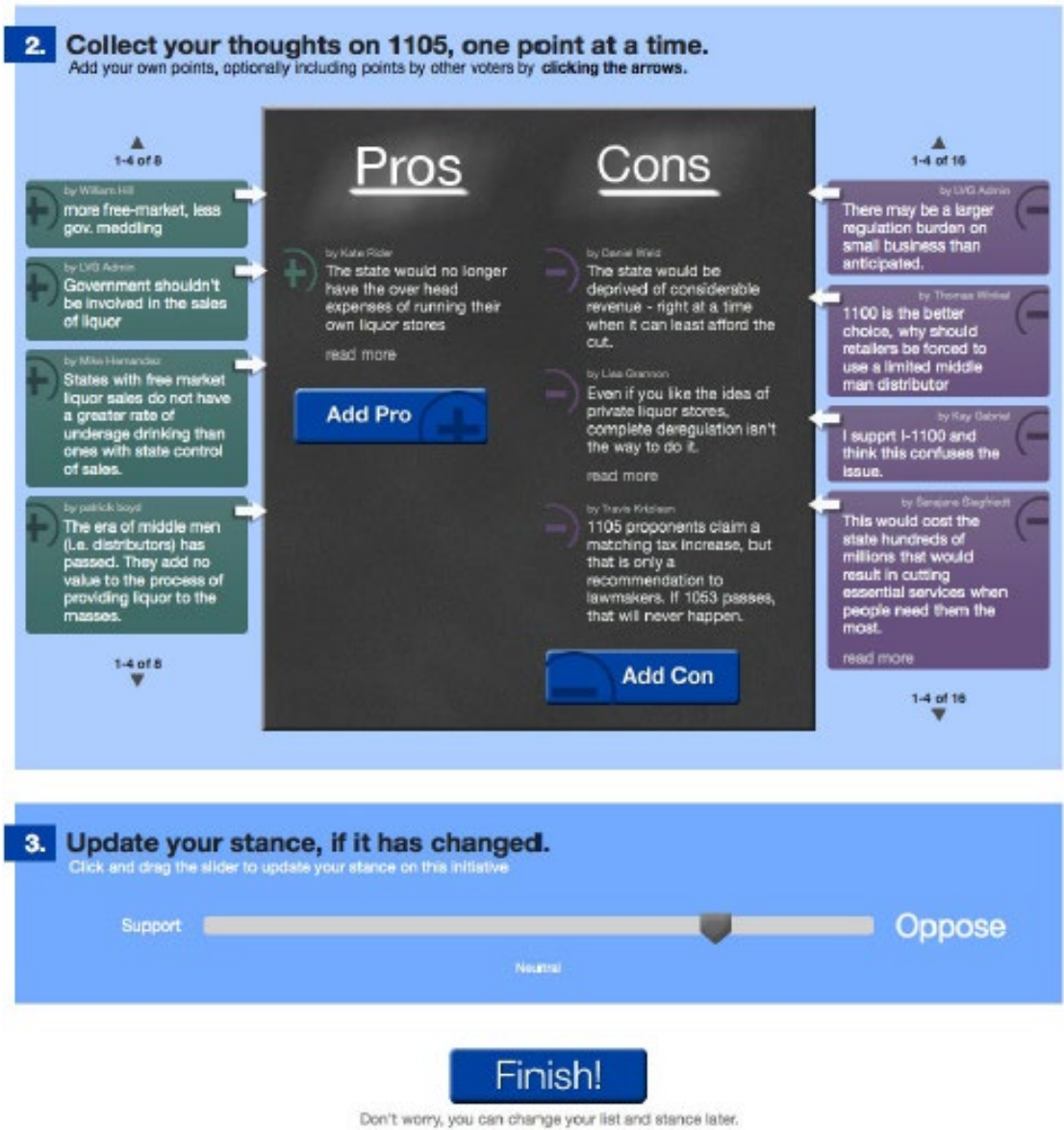

After inputting one's own views after weighing the pros and cons, the following interface shows the user how the rest of users position on the spectrum of views and allows them to share their position on social media.

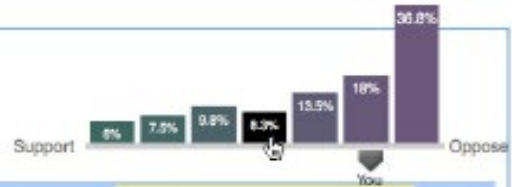

ConsiderIt Washington implemented a post-election survey with 52 respondents and a user study of 7 participants to evaluate their platform. The average age of survey respondents was 42.5 (sd = 17.1), 50.3% male, overwhelmingly white, and high educational attainment (25% with postgraduate degrees). According to survey respondents, "(s)eeing opposing points was seen as the most helpful (mean = 3.30, sd = 1.23), possibly because they serve as a check on one's own decision." "46.3% of survey respondents claimed that they actually changed their stances on at least one measure."

In summary, ConsiderIt is a public platform that is open to all Washington State residents. In order to encourage participation, the project team worked with local newspapers, schools, and social organizations to send the message out. The numbers of registered users and inputs were similar to COLLAGREE and ODSG. As neither random sampling nor user survey is successful in the project, the representativeness of the users is likely to be skewed, or at least unknown. In terms of the outputs, the interface design allows to show the users' overall opinion distribution on the measures, which contributes to the design of a Voter's Guide. However, it is also unknown whether these aggregated results of citizens' opinions and the Voter's Guide affect any voting results or policy making later on.

Key Takeaways

| DOs | Do Nots |
| --- | --- |
| - Partner with local newspapers, organizations, and schools to recruit participants | - Don't expect that citizens will participate if they know there is a platform |
| - Pros and cons interface helps participants to reflect on their own opinions | - Don't just let participants see each other's views through the interface. Create direct interactions. |
| - Set a tangible out comes (e.g., the Voter's Guide) so participants' contribution is meaningful | - Don't forget to evaluate citizens' experience using data-driven methods |

*Case 4: Decidim from Spain*

Decidim stands for "let's decide" in Catalan. Spain has been leading in democratic innovations since the Indignadas Movement in 2011 and the wins of the new Left-wing parties in 2015 elections (Bravo et al.,2019). Decide Madrid, a digital participatory platform launched by the Madrid City Council, began to experiment with various participatory processes in 2015[2]. The platform was based on the Consul software, which provided an initial prototype to Decidim. Decidim Barcelona was the first instance of Decidim and launched in 2016. The Decidim platform served as not only an online discussion space but also an event management system to schedule offline meetings. Thanks to its open source nature, Decidim was later adopted by other Spanish municipalities and local governments in other countries such as Japan[3]. The platform can be replicated in as many instances as needed, with one single installation. The maintenance and updates can be done by one single entity. This is particularly friendly to medium-level and small local governments when their technical capacity is limited.

The platform has an interface that allows the users to initiate different types of citizen participation, or what Decidim calls "participation spaces". Governments can choose the type of participation space that suits their needs the best. Under each type of participation spaces, governments can mix multiple participation components such as in-person meetings, surveys, proposals, debates, results and comments. It is particularly mentionable that there is the accountability component, which can be activated to monitor the degree of execution of the proposals, and people can comment on it. Ordinary citizens can verify their identities so they can vote on the proposals to participate in decision-making.

---

[2] The materials here are mostly adapted from Decidim's own documentation. See link here: https://docs.decidim.org/en/develop/understand/about#_independence_empowerment_and_affiliation
[3] Meta Decidim Japan. See link here: https://meta.diycities.jp/

Decidim Barcelona lasted for 2 months from February 1 and mid-April 2016. The platform attracted 25,435 online participants and received 10,860 proposals (9,560 initiated by citizens) and 18,192 comments (Bravo et al., 2019). The platform is open for the public so anyone who knows about this platform can register and participate. Verified users who Have provided their postal address and ID card can vote on the proposals. There was no random sampling of the population and the participation was based on self-selection. After logging into the platform, there were no training information on the issue provided to citizens either. "The local government committed itself publicly to accept proposals that received the most votes, but the most commented ones were also considered. In total, 75% of the proposals presented for the Strategic City Planning were accepted... However, the local government reserved the right to filter and reject proposals that were not in line with their political priorities." (Bravo et al., 2019, p5684) Zooming onto one particular topic about tourism apartment, Bravo and colleagues (2019) found that the majority of proposals (51.7%) on the platform have not generated any debate at all; The debate has mostly been dominated by few users who have posted the majority of comments. But the good news is that the levels of reciprocity and reflexivity shown in the posts are not low at all, although the deliberation quality tends to drop over time.

In summary, Decidim Barcelona is a platform that is open to all Barcelona residents. As the platform is commissioned by the Barcelona government, the commitment from the government side is clear and high. This high commitment to adopt citizen proposals with highest votes or comments greatly encourages interested citizens to join. The number of registered users is much higher than other platforms reviewed here. However, due to self-selection, the representativeness of citizens is unknown. Due to lack of training information and efficient moderation, it is also unknown how much this platform facilitates deliberation and moreover, changes citizens' opinions towards better arguments.

Key Takeaways

| DOs | Do Nots |
|---|---|
| - Make government commit publicly | - Don't expect that citizens will participate if there is a platform |
| - Use open source software to build the platform | - Don't expect citizens are equipped with necessary information to participate |
| - Combine both online and offline participation | - Don't underestimate the necessity of moderation |

*Case 5: Global citizen assembly*

Citizens' Assembly (CA), or Citizens' Jury, can be dated back as early as the 1970s in the US and Germany. Its recent revival was seen in the UK, Iceland, Canada, Netherlands, and Poland. "A citizens' assembly is a group of people from different walks of life, who come together to learn about a certain topic, to deliberate on possible action, make proposals to governments and leaders and generate ideas to galvanise wider change. Members of a citizens' assembly represent a miniature version of the place in question (say, a country or city, or in this case the world), based on demographic criteria such as gender, age, income and education level." (Global Assembly Information Booklet, 2021) Most CAs were run offline among small groups of citizens. Recently, at least one UK city and one French city decided to make CA a permanent policy making institution, allowing 100 ordinary citizens to input in policy making for about one year. Global Assembly (GA) is a global expansion of CA and focuses on the climate change issue. A group of scholars from University of Canberra, Australia initiated a CA on genome editing with an Australian research fund. The group then worked with foundations, local NGOs, academic partners, etc to take CA to GA. The GA was completed in 2021 and its outcomes were presented at COP26, as voices from the global citizens on important global issues such as climate change. A report is upcoming soon.

GA claims to have an advanced sampling strategy that is based on a fair algorithm (Flanigan et al., 2021). The following graph illustrates this sampling process. The selection started from choosing 100 participants from multiple locations around the globe. Local community hosts were first recruited to gather a pool of participants for their locations. Then the project team randomly selected participants from the pool. The participants were 50% women, 10% no formal education, 70% living on $10 or less a day, 17% from India, 13% white.  The numbers show that there is a high level of diversity among the participants. But it is still far from claiming that this group of participants is representative of the world population.

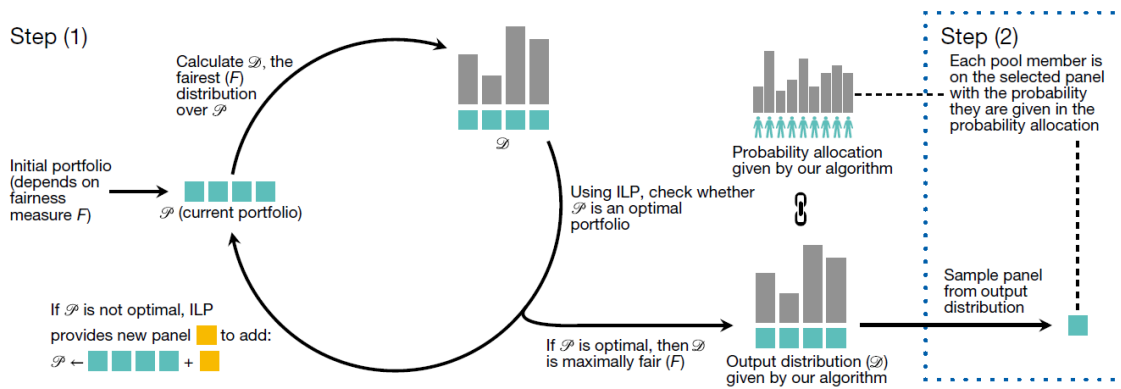

An open source online meeting platform called Jitsi was used in the synchronous online meetings. Details about the platform are not fully released but from the tweets the project posted, there were a group translators, note takers, and facilitators helping with the live discussion. Participants spent 68 hours over 11 weeks to learn about the issue and then deliberate among themselves. In the meantime, local community hosts organized their own local assemblies, although their discussion was not directly input into the Core Assembly's declaration.

In summary, GA is one of the first ambitious attempt to bring a global citizen assembly to the table of global policy making. Although it is too early to tell what actually works and what does not for this GA, the global scope of GA is a clear feature that distinguishes itself from other more local or national level deliberation projects. The visibility of GA, compared to other deliberation projects, is also higher. The presentations GA makes to COP26 and the documentary videos GA makes and releases all help to bring such visibility to GA. GA does deliver a Declaration to the world leaders at COP26, although how much this Declaration influences global policy making and how likely GA will be institutionalized in global policy making are still unknown.

Key Takeaways

| DOs | Do Nots |
| --- | --- |
| - Partner with global NGOs, academics, political leaders as well as local communities | - Don't expect that policy makers will listen if citizens make some decisions |
| - Select participants at the global scale with carefully designed sampling methods | - Don't expect that advanced sampling strategy will guarantee a representative sample of the world population |
| - Raise visibility by using creative media such as documentaries | - Don't underestimate possible backlashes when visibility is high |

**Towards Hybrid Deliberation**

"Offline workshops are better suited for reaching stakeholders and having focused discussions. On the other hand, online workshops can incorporate input from a diverse group of people, such as the young generation or busy business people." Hal Seki, Code for Japan, 2022[4].

"When possible, digital tools should be chosen alongside in-person methods." OECD, 2022.

The report ends with a recommendation to use hybrid deliberation, and being hybrid in three senses: (1) hybrid in online and offline: while the online platform is better for reaching the mainstream population, the face-to-face method suits the elder, less-educated, and lower-income citizens more; (2) hybrid in citizen- and government-led: while researchers can help design, recruit, and moderate the deliberation, government leadership is necessary to mobilize citizen participation and bring the deliberation results into implementation; (3) hybrid in both democratic and non-democratic political systems: although there are contentions on whether deliberation serves the purpose of democracy in non-democratic systems, practicing deliberation is still better than not practicing it.

While transferring these experiences to EU, we need to pay attention to the following issues: (1) the internet infrastructure in the EU location: this will determine how much online deliberation can reach the majority of the population and which minority groups need to be purposely recruited through offline deliberation; (2) the citizen-government relationship in the EU location: depending on how much the government trusts citizens or civil society organizations and vice versa, the hybrid deliberation can be led by either citizens or government officials; (3) the political system and culture in the EU location: depending on how the political institution works, the hybrid deliberation can insert itself into the institution at the right moments, ranging from laws and regulations to political leaders' election promises.

---

[4] "Lessons from Deliberative Democracy using Decidim", accessed at https://youtu.be/Kl9mJc-Vfqc